\documentclass{aa}  
\usepackage{floatflt,graphicx}
\usepackage{txfonts}
\def\deg{$^{\circ}$}

\begin{document}

%
%

\title{Charge-Transfer induced EUV and Soft X-ray emissions in the Heliosphere}
%
%
\author{D.Koutroumpa\inst{1} \and R.Lallement\inst{1} \and V.Kharchenko\inst{2} 
\and A.Dalgarno\inst{2} \and R.Pepino\inst{2} \and V.Izmodenov\inst{3} \and E.Qu\'emerais\inst{1}}
%
\offprints{Dimitra Koutroumpa, \email{dimitra.koutroumpa@aerov.jussieu.fr}}
\institute{Service d'A\'eronomie du CNRS, 91371, Verri\`eres-le-Buisson, France
\and Harvard-Smithsonian Center for Astrophysics, 60 Garden Street, Cambridge, MA 02138
\and Lomonosov Moscow State University, Department of Aeromechanics and Gas Dynamics, Faculty of Mechanics
and Mathematics, Moscow 119899, Russia}
\date{Received 23 Mars 2006/ Accepted 28 July 2006}
%
%
\abstract{}
{We study the EUV/soft X-ray emission generated by charge transfer between solar wind heavy ions and interstellar neutral atoms and 
variations of the X-ray intensities and spectra with the line of sight direction, the observer location, the solar cycle phase and the 
solar wind anisotropies, and a temporary enhancement of the solar wind similar to the event observed by Snowden et al. (2004) during the XMM-Hubble 
Deep Field North exposure.}
{Using recent observations of the neutral atoms combined with updated cross-sections and cascading photon spectra we have computed self-consistent 
distributions of interstellar hydrogen, helium and highly charged solar wind ions for a stationary solar wind and we have constructed monochromatic 
emission maps and spectra. 
We have evaluated separately the contribution of the heliosheath and heliotail, and included X-ray emission of the excited solar wind 
ions produced in sequential collisions to the signal.}
{In most practicable observations, the low and medium latitude X-ray emission is significantly higher at minimum activity than at 
maximum, especially around December. This occurs due to a strong depletion of neutrals during the high activity phase, which is not compensated 
by an increase of the solar wind flux. For high latitudes the emission depends on the ion species in a complex way. Intensity maps are in general 
significantly different for observations separated by six-month intervals. Secondary ions are found to make a negligible contribution to the X-ray 
line of sight intensities, because their density becomes significant only at large distances. The contribution of the heliosheath-heliotail is always 
smaller than 5\%. 

We can reproduce both the intensity range and the temporal variation of the XMM-HDFN emission lines in the 0.52-0.75 keV interval, using a simple 
enhanced solar wind spiral stream. This suggests a dominant heliospheric origin for these lines, before, during and also after the event.}{}
\keywords{Heliosphere -- solar wind -- X-rays}
\titlerunning{HELIOSPHERIC X-RAY EMISSION}
\authorrunning{KOUTROUMPA ET AL.}
\maketitle

\section{Introduction}
That objects as cold as comets emit copious X-rays was a surprising discovery (Lisse et al, 1996). The emission mechanism, 
first proposed by Cravens (1997), is now convincingly demonstrated as a major source of cometary and planetary X-rays. Charge exchange (CX) 
collisions of highly charged ions of the solar wind (SW) with the neutral atoms and molecules from the coma produce cascades of photons in the 
extreme ultraviolet (EUV) and soft X-ray regions (H\"{a}berli et al. 1997, Krasnopolsky, 1997, Kharchenko and Dalgarno, 2000, Schwadron and Cravens, 2002). 

Cox (1998) pointed out that neutral interstellar (hereafter IS) and geocoronal atoms would also undergo CX with the SW ions, and generate soft X-rays 
throughout the heliosphere. The heliospheric  emission has been subsequently modelled by 
Cravens et al (2001) and Robertson et al (2001), who explained satisfactorily the global correlation observed by Freyberg (1994, 1998) between  the 
so-called Long Term Enhancements (LTE's) measured by the ROSAT satellite  and strong solar wind events. The geocoronal emission has 
been investigated by Robertson and Cravens (2003) and measured during Chandra observations of the dark moon (Wargelin et al, 2004). Earlier, Gruntman 
(1998) modelled the EUV emission produced by the recombination of alpha particles of the SW colliding with interstellar atoms, and showed 
the substantial influence of the solar wind characteristics on the emission.

The intensity of the soft X-rays in the heliosphere was first estimated by Cravens (2000), who found it to be of the same order as the 
soft X-ray emission from the so-called Local Bubble (hereafter LB), the 100 parsecs wide cavity surrounding the Sun, filled with tenuous hot (10$^6$ K) gas 
(Snowden et al, 1994, 1999). More generally, for faint and diffuse astronomical X-ray sources the CX heliospheric emission may 
contaminate significantly the X-ray spectra below 1.5 keV as has been demonstrated by a long duration XMM-Newton exposure 
towards the Hubble Deep Field-North (Snowden et al, 2004, hereafter SCK04).

The cosmic diffuse X- ray emission is the sum of extragalactic, halo and LB contributions (Kuntz and Snowden, 2000). 
It is difficult to separate out the CX emissions from those of the Local Bubble (LB). Shadowing is of limited use because of the absence of 
clouds of sufficient density and emission lines dominate the continuum in the CX spectrum and the LB spectrum.

As seen from the Sun, heliospheric CX emission maxima will be aligned along the interstellar wind axis (Cox, 1998). The ROSAT 6-month 
survey necessary to reconstruct full sky soft X-ray maps did not show this alignment so that low upper limits to the local emission were derived. 
However, it has been demonstrated that parallax effects connected with the ROSAT all-sky survey geometry destroy the axial symmetry and cause a much 
smoother emission pattern, so that substantial contamination 
by CX is not excluded (Lallement, 2004).

Robertson and Cravens (2003a,b) constructed sky maps of the heliospheric and geocoronal CX intensity, using global efficiency factors 
representing all ions at once. Pepino et al. (2004) carried out similar calculations of the spectra emitted by the individual ions and of the 
corresponding power densities taking separate account of collisions of hydrogen and helium and distinguishing between the fast and slow solar winds.

We extend the study to the calculation of line of sight (LOS) spectral emission maps. We explore potential sources of variability of the received 
signals and quantify the resulting intensities. We take into account the observer location, the solar cycle phase, the LOS and we calculate the intensities 
of the strongest emission lines and the contributions of the outer heliosphere. We investigate the contribution of secondary ions created sequentially 
by CX. We examine the effects of temporary solar wind enhancements and solar rotation (Cravens 2000) using a realistic model 
of the expanding solar wind. Observations of the soft diffuse X-ray background with the XMM - Newton telescope are analysed.

In section 2 we describe the model developed for the calculation of the X-ray emission. In section 3, we present the heliospheric X-ray emission maps 
derived from the model and discuss the effects of the solar cycle, observing location on the emission pattern. In 
section \ref{secCE_section}\ we evaluate the contribution of the secondary ions and in \ref{outerHelio_section} the contributions of the heliosheath 
and heliotail. In section \ref{helio_spectra}, we present X-ray and EUV spectra for lines of sight at different helio-ecliptic latitudes, corresponding to slow and fast 
solar winds, and for neutral gas distributions that depend on solar activity conditions. In section \ref{Snow_mod} we simulate the temporal variations of 
the heliospheric X-ray emission generated by a solar impulsive event for solar wind conditions and a geometry corresponding to the XMM-Newton 
Hubble Deep Field-North (HDFN) exposure of June 1, 2001. The model results are compared with the XMM-Newton observations of the diffuse background 
spectra by SCK04. Finally, in the last section \ref{conclusions}, we summarize and discuss the results. 

\section{Description of the model}
Our simulation of the heliospheric CX emission has four steps. The first is the computation of the density distribution of IS H and He atoms 
in response to the solar wind and solar EUV conditions for solar minimum and maximum activity. In the second step, these density 
distributions are used to calculate densities of heavy solar wind ions (X$^{Q+}$), modulated by collisions with the neutral heliospheric gas. In the 
third step we use the self-consistent density grids of H and He neutral atoms and solar wind ions to calculate the X-ray emissivity due 
to the CX collisions. Finally these emissivities are used to calculate the total intensity along all lines of sight.

\subsection{Hydrogen and Helium Density grids}

In order to calculate the neutral H density distribution we have adapted a so-called classical $^{\prime}$hot model$^{\prime}$ used in computing the 
interplanetary-interstellar H Lyman-$\alpha$ emission (Lallement et al. 1985a, b). The IS hydrogen flow, after 
crossing the heliospheric interface, is described as a single Maxwellian flow. The density distribution reflects the action of gravity, 
radiation pressure and losses due to solar wind CX and solar EUV ionization and recognises the latitudinal 
anisotropy of the loss terms. Although the CX with H$^{+}$ is not an H loss process, the newly created H atoms are expanding radially 
at a high velocity and their contribution to the density is negligible. Thus, we treat resonant CX as if it were a pure ionization process of 
the IS hydrogen gas. The parameters specifying the IS neutral hydrogen are: $n_H$(at 100 AU) = 0.1 cm$^{-3}$, $T$ = 13000 K, $V_o$ = 21 km/s, 
$\lambda_{UW}$ = 252.3\deg , $\beta_{UW}$ = 8.5\deg (Lallement et al., 2005), where $\lambda_{UW}$ and $\beta_{UW}$ are the helio-ecliptic 
longitude ($\lambda$) and latitude ($\beta$) respectively for the upwind direction of the incoming neutral H flow. We employ 
a density grid $n_H[r, \lambda, \beta]$ with a 1\deg x 1\deg resolution for $\lambda, \beta$ and for r a variable step increasing with 
distance r from the Sun, from $\delta r$= 0.3 AU at the earth orbit up to $\delta r$= 9 AU at 100 AU.

Helium density distributions in the inner heliosphere have been computed with the model developed by Lallement et al. (2004). The helium distributions 
are described by the classical kinetic model, with the parameters: $n_{He}$(at 100 AU) = 0.015 cm$^{-3}$, $T$ = 6300 K, $V_o$ = 26.2 km/s, 
$\lambda_{DW}$ = 74.7\deg , $\beta_{DW}$ = -5.3\deg  (Witte, 2004, Vallerga et al., 2004, Gloeckler et al., 2004). Resolution in $\theta$ is 1\deg . 
Distance steps are varying logarithmically in the sunward direction and they are very small near the Sun, where the helium is still dense and the 
emissivity due to He is high and varying rapidly.

Solar activity influences differently the heliospheric distributions of H and He atoms. The ratio, $\mu$, of radiation pressure to gravity for neutral 
hydrogen, varies from 0.9 at solar minimum to 1.5 at solar maximum (Woods et al., 2000). For H, the major source of ionization is CX of H with 
solar wind protons, followed by EUV photo-ionization. Ionization rates $\beta_i$ as a function of heliographic latitude are derived from the SOHO-SWAN preliminary data analysis (Qu\'emerais et al., 2006). During 
solar minimum $\beta_i$ is 6.6 10$^{-7}$ s$^{-1}$ at heliographic latitudes between $\pm$ 20\deg and 4. 10$^{-7}$ s$^{-1}$ at 
latitudes over $\pm$ 20\deg and up to the poles. During solar maximum, the anisotropy is less important. It is scaled every 10\deg with $\beta_i$ 
values between 8.4 10$^{-7}$ s$^{-1}$ at the equator and 6.7 10$^{-7}$ s$^{-1}$ at the poles.  

For helium atoms $\mu$ = 0. Radiation pressure is negligible compared to solar gravitational attraction and the 
atoms are gravitationally focused downwind. The main cause of ionization is solar EUV radiation and electron impacts. The mean 
lifetime (the inverse of $\beta_i$) at 1 AU varies from 1.4 10$^7$ s at solar minimum to 0.62 10$^7$ s at 
solar maximum, in agreement with McMullin et al., 2004. 

Recent work has shown strong evidence for an anisotropic distribution of the He 30.4 nm solar irradiance and thus of the He photo-ionization 
rate (Witte et al., 2004, Auch\`ere et al., 2005) and it has been shown that the electron impact ionization rate is also anisotropic 
(McMullin et al., 2004). However, since our model does not include any latitudinal dependence of the helium ionization rate, we use isotropic 
photo-ionization and electron impact ionization.

The radial dependence of electron impact ionization is taken from Rucinski \& Fahr (1989). Lallement et al. (2004) used it in an analysis of 
SOHO-UVCS 58.4 nm data, and found that the rate is appropriate for solar minimum, but requires a three-fold increase for solar 
maximum. In order to have self-consistency we use both the Rucinski \& Fahr (1989) radial dependence and the Lallement et al. (2004) electron 
impact ionization. Further work is, however, certainly needed to improve the accuracy of the helium ionization rate of helium 
(McMullin et al., 2004). 

\subsection{Heavy Solar Wind Ion Distributions}

In the inner heliosphere and in the absence of charge transfer, the density of heavy ions would follow a r$^{-2}$ dependence. Here, 
in addition to the radial expansion, we consider the effect of collisions with interplanetary neutrals. The next step in our model is the 
computation of the heavy ion losses due to CX with hydrogen and helium atoms and the determination of a density distribution for 
each ion species. The process may be written:
\begin{equation}
X^{Q+} \, + \, [H, He]\,  \rightarrow \, X^{\ast (Q-1)+ } \, + \, [H^+, He^+] \,
\end{equation}
and it represents the loss of ion X$^{Q+}$ and the production of ion X$^{(Q-1)+}$. The density of ion X$^{Q+}$ is given by the equation:
\begin{eqnarray}
{\frac{dN_{X^{Q+}}}{dx}} &=& {- \, N_{X^{Q+}}(\sigma_{(H, X^{Q+})} \, n_H(x) + \sigma_{(He, X^{Q+})} \, n_{He}(x))} \nonumber \\
&& {+ \, N_{X^{(Q+1)+}}(\sigma_{(H, X^{(Q+1)+})} \, n_H(x) + \sigma_{(He, X^{(Q+1)+})} \, n_{He}(x))} \label{eqn:secondary}
\end{eqnarray}
where x is the distance along the SW stream lines, $\sigma_{(H,X^{Q+})}$ and $\sigma_{(He,X^{Q+})}$ are 
the hydrogen and helium CX cross-sections and $n_H(x)$ and $n_{He}(x)$ are the hydrogen and helium densities respectively. The first term 
represents the loss due to CX between X$^{Q+}$ and H and He while the second term is the source term due to the equivalent CX process for X$^{(Q+1)+}$. 
For bare ions (C$^{6+}$, N$^{7+}$, O$^{8+}$) the source term is, of course, absent, and in a first approximation we neglect it also for the lower state ions. 
We will discuss its role and justify this assumption in section \ref{secCE_section}. 

If we neglect this term, the ion density as a function of distance from the Sun is described by the equation:
\begin{equation}
N_{X^{Q+}}(r) = \frac{N_{X^{Q+}o}}{r^2}  exp(-\int_{r_o}^{r} (\sigma_{(H, X^{Q+})} \, n_H(x) + \sigma_{(He, X^{Q+})} \, n_{He}(x)) dx)
\label{eqn:primary}
\end{equation}
where r is the radial distance from the Sun and the density of ion $X^{Q+}$ at 1 AU is expressed in the form 
$N_{X^{Q+}o}=[\frac{X^{Q+}}{O}]\,[\frac{O}{H^+}]\,n_{H^+o}$, $[O]$ being the total oxygen ion content of the solar wind and $n_{H^+o}$ 
the proton density at 1 AU. The adopted values of 
$[\frac{X^{Q+}}{O}]$ and $[\frac{O}{H^+}]$ for the fast and slow solar winds are given in Table \ref{SWparam}.

We assume that the heavy ion flux is propagating radially from the Sun. For each grid cell (r, $\lambda$, $\beta$), where $\lambda$ and $\beta$ 
are the ecliptic coordinates, varying 1\deg by 1\deg , the corresponding heliographic coordinates $\lambda_{Helio}$ and $\beta_{Helio}$ are 
calculated, and $\beta_{Helio}$ is used to discriminate between slow and fast solar wind. 

During solar minimum we adopt the Slow Solar Wind composition (SSW) at an average velocity of 400 km/s, accelerated in the equatorial zone 
($\beta_{Helio}$ = [-20\deg , 20\deg ] ), and the Fast Solar Wind (FSW) composition with an average velocity of 750 km/s, accelerated at higher 
heliographic latitudes ($\beta_{Helio}$ = [-20\deg , -90\deg ], [20\deg , 90\deg ]). The proton density $n_{H^+o}$ is 6.5 cm$^{-3}$ 
for the SSW and 3.2 cm$^{-3}$ for the FSW.  During solar maximum we assume that all the solar wind is in the SSW state. These 
values are consistent with the ionization rates we have used for the computation of the neutral densities. In Table \ref{SWparam} we summarize 
the parameters characterizing the two different states of the Solar Wind, as well as abundances $[\frac{X^{Q+}}{O}]$ and CX cross sections 
with H and He for each heavy ion species (and charge state) used in our analysis. The charge and elemental abundances of heavy ions in the slow and 
fast solar wind are adopted from Schwadron \&\ Cravens, (2000). We have updated values of cross sections of ion collisions with H and He atoms. The 
selective and total cross sections of the charge transfer collisions of the heavy solar wind ions have been constructed using relevant data from 
the laboratory measurements and theoretical calculations reported in table \ref{SWparam} .

\begin{table*}
\caption{Slow and Fast Solar Wind Parameters. Cross sections for CX between SW heavy ions and H and He are based on theoretical 
and experimental work of [Beijers et al (1994), Bliman et al (1992), Bonnet et al (1985), Dijkkamp et al (1985a,b), Fritsch et al (1996), 
Greenwood et al (2001), Harel et al (1992, 1998), Ishii et al (2004), Iwai et al (1982), Lee et al (2004), Liu et al (2005), Phaneuf et al (1987), 
Richter et al (1993), Shimakura et al (1992), Suraud et al (1991), Wu et al (1994)].\protect \newline
$^{a}$ Including most important lines, $^{b}$ Schwadron \&\ Cravens (2000), $^{c}$ Rough estimate from ACE data.\label{SWparam}}
\begin{flushleft}
\begin{minipage}[t]{\linewidth}
   \renewcommand{\footnoterule}{}
\begin{tabular}{l|l|ccc|ccc}\hline
\multicolumn{2}{l}{SW Type}  & \multicolumn{3}{|c|}{Slow} & \multicolumn{3}{c}{Fast}\\ \hline 
\multicolumn{2}{l}{$V_{SW} (km/s)$}  & \multicolumn{3}{|c|}{400} & \multicolumn{3}{c}{750}\\
\multicolumn{2}{l}{$[\frac{O}{H^+}]$}  & \multicolumn{3}{|c|}{1/1780} & \multicolumn{3}{c}{1/1550}\\
\multicolumn{2}{l}{$n_{H^+o}$ at 1 AU(cm$^{-3}$)}  & \multicolumn{3}{|c|}{6.5} & \multicolumn{3}{c}{3.2}\\ \hline
 &Spectral Lines (keV) $^{a}$ & & $\sigma_{(H, X^{Q+})}$ & $\sigma_{(He, X^{Q+})}$ &  & $\sigma_{(H, X^{Q+})}$ & 
$\sigma_{(He, X^{Q+})}$ \\
$X^{Q+}$ & Produced by $X^{*(Q-1)+}$& $\,[\frac{X^{Q+}}{O}]\, ^{b}$ &  
$(10^{-15}\,cm^2)$ & $(10^{-15}\,cm^2)$ & $\,[\frac{X^{Q+}}{O}]\,^{b}$ & $(10^{-15}\,cm^2)$ & $(10^{-15}\,cm^2)$\\ \hline
C$^{6+}$&(0.37, 0.44, 0.46) & 0.318 & 4.16 & 1.50 & 0.085 & 4.63 & 1.50\\
C$^{5+}$&(0.3, 0.35, 0.37) & 0.210 & 2.00 & 1.40 & 0.440 & 2.90 & 1.11\\
N$^{7+}$&(0.25, 0.5, 0.6, 0.62, 0.64) & 0.006 & 5.67 & 2.00 & 0.000 & 5.55 & 2.00\\
N$^{6+}$&(0.42, 0.43, 0.5)& 0.058 & 3.71 & 1.26 & 0.011 & 3.32 & 1.49\\
N$^{5+}$&(0.05, 0.059, 0.065)  & 0.065 & 2.27 & 1.41 & 0.127 & 2.92 & 1.07\\
O$^{8+}$&(0.33, 0.65, 0.77, 0.82, 0.84) & 0.070 & 5.65 & 2.80 & 0.000 & 6.16 & 2.80\\
O$^{7+}$&(0.561, 0.569, 0.574) & 0.200 & 3.40 & 1.80 & 0.030 & 3.70 & 1.97\\
O$^{6+}$&(0.072, 0.083, 0.094, 0.107)  & 0.730 & 3.67 & 0.96 & 0.970 & 3.91 & 1.31\\
Ne$^{9+}$&(0.905, 0.915, 0.922) & 0.030 $^{c}$ & 7.20 & 2.40 & 0.006 $^{c}$ & 7.20 & 2.40\\
Ne$^{8+}$&(0.126, 0.141, 0.187)  & 0.084 & 3.70 & 1.30 & 0.102 & 3.00 & 1.10\\
Mg$^{11+}$&(1.33, 1.34, 1.35, 1.37) & 0.035 $^{c}$ & 7.5 & 2.6 & 0.001 $^{c}$ & 7.5 & 2.6\\
Mg$^{10+}$&(0.28, 0.29, 0.3, 0.31) & 0.098 & 3.73 & 1.00 & 0.029 & 2.50 & 0.90\\ \hline
\end{tabular}
\end{minipage}
\end{flushleft}
\end{table*}

We have calculated the number of X$^{Q+}$ ions lost due to CX collisions with H and He atoms for the SW plasma flux propagating in the radial 
direction. For that we use a refined grid such that each grid cell is smaller than the H and He grid cells at each distance from the Sun so 
that we can obtain the best resolution for the affected regions of both neutral species.

\subsection{Calculating the X-Ray Emission}
The final step of our simulation is the computation of the directional intensity I$(\lambda,\beta)$ of X-ray emission lines, resulting from 
CX collisions of the heavy SW ions X$^{Q+}$ with the heliospheric H and He atoms. We compute the line of sight intensity I, 
seen by an observer at Earth orbit at different dates. CX collisions, producing X-ray photons, are described by 
the equation:
\begin{eqnarray}\label{eqn:yld}
{X^{Q+} \, + \,  [H, He]} & \rightarrow & {X^{*(Q-1)+} \, + \, [H^+, He^+]}  \\ 
 & \rightarrow & {X^{(Q-1)+} \, + \, [H^+, He^+] \, + \, [Y_{(E_i,H)}, Y_{(E_i,He)}]}\nonumber
\end{eqnarray}
where $Y_{(E_i,H)}$, $Y_{(E_i,He)}$ is the photon yield for the spectral line $E_i$ induced in the CX between the ion $X^{Q+}$ with 
H and He respectively. The values of quantum yields in collisions of the most important solar wind ions have been presented in articles on the 
charge transfer mechanism of cometary X-ray emission [Kharchenko and Dalgarno (2000, 2001); Rigazio et al. (2002); Kharchenko et al. (2003); 
Pepino et al. (2004)]. The photon energies E$_i$ and relative intensities of different emission lines of the CX spectra have been computed for the fast 
and slow winds. 

The volume collision frequency R$_{X^{Q+}}$ in units of cm$^{-3}$ s$^{-1}$ of $X^{Q+}$  ions with the neutral heliospheric atoms is 
given by the equation:
\begin{eqnarray}\label{eqn:nbc}
{R_{X^{Q+}}\,(r)} &=& {N_{X^{Q+}o}(r)\,v_{rel}\,(\,\sigma_{(H, X^{Q+})} \, n_H(r) + \sigma_{(He, X^{Q+})} \, n_{He}(r)\,)} \nonumber\\  
 &=& {R_{(X^{Q+},H)}\,(r) + R_{(X^{Q+},He)}\,(r)} 
\end{eqnarray}

If we consider only the region inside the termination shock, where solar wind ions are supersonic, then the relative velocity between the 
solar wind ions and the IS neutrals, $\vec {v_{rel}} = \vec {V_{SW}} - \vec {v_n} $ in equation \ref{eqn:nbc}, can be approximated by $V_{SW}$, since 
the neutral velocity $v_n \ll V_{SW}$. 
With the exception of section \ref{outerHelio_section} all the results have been obtained with this assumption. In section \ref{outerHelio_section} 
we examine in detail the case of the outer Heliosphere (the heliosheath and heliotail) where the IS neutral velocity and ion thermal velocity are 
no longer negligible. 

Thus, for the inner heliosphere, the number of $h\nu$-photons emitted per second from unit volume is given by the formula:
\begin{equation}
\varepsilon_{h\nu}\,(r) = R_{(X^{Q+},H)}(r)\,Y_{(h\nu ,H)} + R_{(X^{Q+},He)}(r)\,Y_{(Eh\nu ,He)}
\end{equation}

The intensity measured at energy $h\nu$ for an observer in position $\vec{O}(\lambda_E,\beta_E)$ for a line of sight $\vec{LOS}(\lambda,\beta)$ is 
\begin{equation}   \label{itot}
I_{h\nu}(\vec{O},\vec{LOS}) = \frac{1}{4\pi}
\int_0^{100 AU} \varepsilon_{h\nu}\,(r)\, ds 
\end{equation}

During solar maximum, the computation of the intensity is quite simple since we consider only one state of the Solar Wind, the slow one. This is not 
the case during solar minimum and especially for LOS towards high latitudes. In Fig.\ref{los_geo} (Supporting Online Material, hereafter SOM) 
we present a simplified observation geometry for such lines of sight for solar minimum conditions. In the figure, we have considered the Solar 
Equator parallel to the ecliptic plane but in reality the two planes have an angle of incidence of about 7.25\deg which is taken into account in 
the modelling. As shown in figure \ref{los_geo} (SOM), a high latitude LOS crosses regions of both slow and fast SW and thus, the total 
intensity is a mixture of SSW and FSW induced  photons. This is taken into account in our simulation when we interpolate in our heavy-ion grid for 
each point on the LOS according to its distance and latitude.

\section{Charge transfer EUV and X-ray maps of the Heliosphere}

\subsection{Emissivity Maps}
Figure \ref{isoemiO5_O6} (SOM) shows the contour maps of the total emissivity of excited O$^{\ast 5+}$ and O$^{\ast 6+}$ ions produced in 
the charge transfer collisions of the SW ions with the hydrogen and helium gas during solar minimum. The top panels present iso-emissivity 
contours at positions (x, y) between $\pm$15 AU in the plane defined by the IS He wind axis and the vector ($\lambda$ = -16\deg , $\beta$ = 0\deg ) 
which is very close to the ecliptic plane and, thus, contains mainly slow SW. In these figures, the IS wind is travelling from left to right. We 
recognize on the upwind side the crescent shape due to the $r^{-2}$ dependence of the solar wind flux and the spatial distribution of hydrogen. 
On the downwind side we find the excess of emissivity due to the focusing of He neutrals in the He cone. 

The bottom panels of fig. \ref{isoemiO5_O6} (SOM) present iso-emissivity contours at positions (x, z) between $\pm$ 15 AU in the plane 
defined by the IS He wind axis ($\lambda$ = 74\deg , $\beta$ = -5\deg ) and the heliospheric polar axis.  The IS wind comes from the left and 
heads on -5\deg downwards with respect to the z axis. The two panels stress the discontinuity at $\pm$ 20\deg between the SSW and the FSW during 
solar minimum, as well as the contrast between the ions characterizing the two SW types. The discontinuity boundaries appear very sharp because we 
consider a simplified anisotropic model for the solar wind. The left panels present iso-emissivity contours for the O$^{5+}$ line at 0.08 keV (sum 
of the lines 0.072 and 0.082 keV), generated from O$^{6+}$, abundant in the fast SW. The right column presents contours for the sum of intensities 
of O$^{*6+}$ emission lines of 0.561, 0.569 and 0.574 keV, induced in radiative transitions from the triplet and singlet excited states of O$^{*6+}$. 
We show in fig. \ref{isoemiO5_O6} (SOM) the average energy of O$^{*6+}$ photons: 0.57 keV. The O$^{7+}$ SW ions, producing 
excited O$^{*6+}$, are more abundant in the SSW.

\subsection{Intensity Maps}
In figures \ref{fullskyDW} (and \ref{fullsky121}, \ref{fullskyCW}, \ref{fullsky211}, SOM) we present full-sky maps of the heliospheric X-ray 
emission in ecliptic coordinates. Color in these figures represents intensities of the CX emission given in 
10$^{-9}$ erg cm$^{-2}$ sr$^{-1}$ s$^{-1}$, the red colour corresponding to minimum and the blue to maximum values. We also mark in each map contours 
at 10\%\ (white), 50\%\ (grey) and 90\%\ (black) of maximum value. We have removed from every map a data portion of 20\deg x 20\deg around the solar 
disk where no instrument can observe. In the following we examine the different effects illustrated in these maps.

\subsubsection{Solar Cycle Effects}
\begin{figure*}
\centering
\begin{minipage}[b]{0.21\textwidth}
\centering
\caption{Solar maximum and solar minimum full sky monochromatic for the O$^{7+}$ line at 0.65 keV and the O$^{5+}$ line at 0.08 keV. 
The observer is situated 
at 75\deg (left column) and at 251\deg (right column) ecliptic longitude. The color scale is in 
units of 10$^{-9}$ erg cm$^{-2}$ sr$^{-1}$ s$^{-1}$, red colour corresponding to minimum and blue to maximum values. 
The masked area is the 20\deg x 20\deg region around the solar disk.
The map is shown in ecliptic coordinates.}
\label{fullskyDW}
\par\vspace{0pt}
\end{minipage}
\begin{minipage}[b]{0.79\textwidth}
\centering
\includegraphics[width=\textwidth ,height=!]{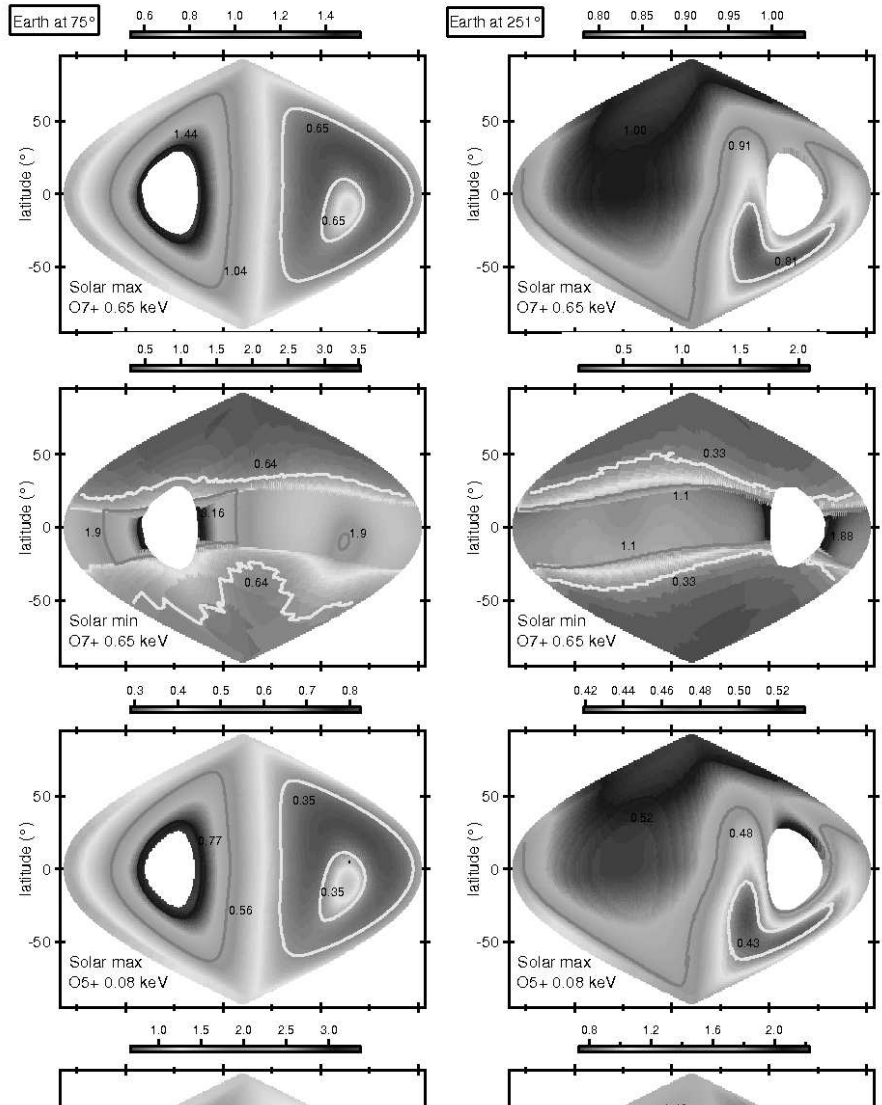}
\par\vspace{0pt}
\end{minipage}
\end{figure*}

Again, in figures \ref{fullskyDW} and \ref{fullsky121}, \ref{fullskyCW}, \ref{fullsky211} (SOM) we chose two oxygen ions characterizing 
the slow or the fast SW, in order to emphasize the contrast between the two SW states. In the two top panels of each column we illustrate the O$^{7+}$  
0.65 keV line for solar maximum and solar minimum conditions. In the third and fourth panel of each column we present the O$^{5+}$ 0.08 keV line for 
solar maximum and solar minimum respectively. 

The striking features are: (i) At solar minimum, the modeled intensity in the equatorial region is conspicuously different from regions at 
medium or high latitudes. This is entirely due to the solar wind latitudinal structures, separating the slow from the fast solar wind, shown in 
figure \ref{los_geo} (SOM). For the fast solar wind, the major contributors to the CX emission are low charge such as 
C$^{4+}$, N$^{5+}$ and in particular O$^{5+}$, illustrated in Figure \ref{fullskyDW} . These ions emit in the EUV (E $\leq$ 150 eV). 
The slow solar wind is characterized by highly charged ions C$^{6+}$, N$^{7+}$, O$^{8+}$ and Ne$^{9+}$ which produce harder X-ray emitting ions, 
as for example O$^{7+}$ illustrated in the maps. 

(ii) The X-ray emission pattern for solar maximum is very similar for both O$^{5+}$ and O$^{7+}$ lines, but for solar minimum there are strong 
differences. While O$^{7+}$ emission is almost absent at high ecliptic latitudes, O$^{5+}$ is much brighter due to the difference 
in relative abundance of O$^{8+}$ and O$^{6+}$. O$^{7+}$ is generated during CX between O$^{8+}$ ion and H and He. O$^{8+}$ is absent from the 
fast solar wind. The O$^{7+}$ emission is never zero, because there are no lines of sight containing only the FSW.

(iii) At low latitudes, near the solar equator, X-ray emissions are more intense for solar minimum than for solar maximum because the neutral 
atom content is higher during solar minimum. During solar minimum, photo-ionization is less efficient and H and He are less readily destroyed by 
photo-ionization. Further, gravitational pressure exceeds radiation pressure and thus neutral H atoms have incoming trajectories and fill the 
ionization cavity.

\subsubsection{Effect of the Observer's position}
Each column represents a different observing position on the Earth's orbit. We cover half an orbit, about every 45\deg starting downwind 
($\lambda_E$ = 75\deg , fig.\ref{fullskyDW}, left) to upwind ($\lambda_E$ = 251\deg , fig.\ref{fullskyDW}, right). When 
examining the different vantage points on the Earth's orbit, it is no longer the solar wind ion distribution that is the source of the X-ray pattern 
differences, but the H and He distributions in interplanetary space. These differences are even more striking in the solar maximum maps, where 
only one solar wind type is present.

When the observer is situated downwind (see fig.\ref{fullskyDW}, left), emission is dominated by CX with He. Maximum 
intensity is observed for a ring of LOS upwind, around the forbidden solar direction. These LOS cross all the ionization cavity, extending to 
several AU, before reaching the first H atoms. Thus the maximum in intensity is due to He concentrated around and 
behind the sun, on the downwind (DW) side. In this region the focusing cone ($\lambda$ = 75\deg , $\beta$ = -5\deg ) produces an excess of CX 
and thus of X-ray intensity. H atoms are absent downwind, where the ionization cavity limits the neutral atom abundance. This is illustrated by the 
minimum of emission in the LOS forming a large ring around the He cone. 

These characteristic features evolve gradually when moving on the Earth's orbit from downwind to upwind (fig. \ref{fullsky121}, 
\ref{fullskyCW}, \ref{fullsky211} - SOM). Maximum emission is still located close around the sun but the emission pattern loses its axisymmetric form. 
The He cone becomes less and less striking, but still distinguishable by an excess of emission when observing a LOS that crosses it. The minimum X-ray 
intensity obtains a large oval-like pattern in LOS pointing downwind, less and less disturbed by the presence of the He cone as we move upwind. 

On an upwind position of the observer (fig.\ref{fullskyDW}, right), the major source is CX with H with maximum X-ray intensity on the 
northern antisolar lines of sight, towards the incoming flow. Those antisolar LOS cross a much smaller portion of the ionization cavity than LOS 
pointing downwind. They therefore see more H atoms approaching, while they miss He atoms which are absent at distances greater than 1 AU 
from the sun. Minimum X-ray intensity on these maps is thus situated in the southern hemisphere towards the Sun (DW) illustrating 
the major loss of H neutrals in the ionization cavity. Even though there is little He on the upwind maps, it is not completely absent and we 
still find a halo of X-ray emission around the Sun.

\subsection{The Secondary Ion Charge transfer Source Term} \label{secCE_section}
\begin{figure}
\centering
\includegraphics[width=0.45\textwidth ,height=!]{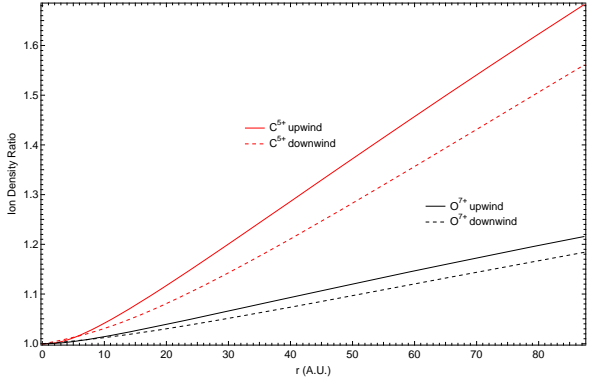}
\caption{Ratio between the total ion density (including secondary CX sources) 
and primary ion density as a function of distance from the Sun.}
\label{rap_sec}
\end{figure}

Secondary ions contribute to the total heliospheric emission. In figure \ref{rap_sec} we show how the ratio of the total ion density, including 
secondary CX-produced ions (eq. \ref{eqn:secondary}), to the density calculated with the primary CX destruction term (eq. \ref{eqn:primary}) only, 
varies with distance to the Sun. We show this ratio for two different ions originating from slow SW regions, C$^{5+}$ and O$^{7+}$, 
moving in two radial directions, and heading upwind or downwind. 

C$^{5+}$ is produced in C$^{6+}$ + H CX collisions with a high value of cross section and depleted in the CX C$^{5+}$ + H and C$^{5+}$ + He collisions, 
with cross sections that are significantly smaller. Moreover, C$^{6+}$ is more abundant than C$^{5+}$ in the slow SW, and CX of C$^{6+}$ 
is a large source of C$^{5+}$. The gain of C$^{5+}$ exceeds the loss and its density increases rapidly at large distances from the Sun. 
For the other ions the source term is less important but it slows the decrease that would occur in its absence. The ion creation term 
becomes effective at large distances from the Sun of 30-40 AU, beyond the region where most of the emission takes place. For an upwind LOS, 
88\% of the total X-ray emission due to H is produced in the first 10 AU and 98\% of the total X-ray emission due to He is 
produced within the first 5 AU We conclude that the secondary source term does not add more than 4\% to the total X-ray emission of  the inner 
Heliosphere, and therefore we can neglect it in a first approach. 

\subsection{The outer Heliosphere Contribution} \label{outerHelio_section}
In the previous sections, we analyzed the various effects of the solar cycle and the observer's position on the total emission maps. The maps 
refer to the inner heliosphere, where the solar wind is supersonic and flows radially. Beyond the termination shock it is forced to 
decelerate and turn backwards at the heliopause. The relative contribution of the inner heliosheath and the heliotail to the total X-ray emission 
has been computed separately using the plasma and neutral distributions of Izmodenov \&\ Alexashov (2003).  In this kinetic-continuum 
model (Baranov and Malama, 1993), the solar wind at the orbit of the Earth is steady and spherically symmetric. The interstellar flow 
which consists of hydrogen atoms is uniform and plane-parallel outside the heliosphere. In these conditions the flow is steady and axisymmetric. 
All charged particles (electrons, protons, alpha-particles, pickup ions) are treated as a single component fluid, described by the hydrodynamic 
Euler equations with source terms that include the effects of CX. The motion of the interstellar atoms in the heliospheric interface is determined 
by solving the kinetic equation. It is assumed in the model that H atoms interact with the plasma by CX only and electron impact is neglected. 
Computations have been carried out far in the tail of the interaction region. This self-consistent neutral-plasma model yields H and ion densities, 
thermal velocities and flow velocities along streamlines. We use these results out to 3000 AU beyond which the production of X rays is negligible.

\begin{figure}
\centering
\includegraphics[width=0.4\textwidth ,height=!]{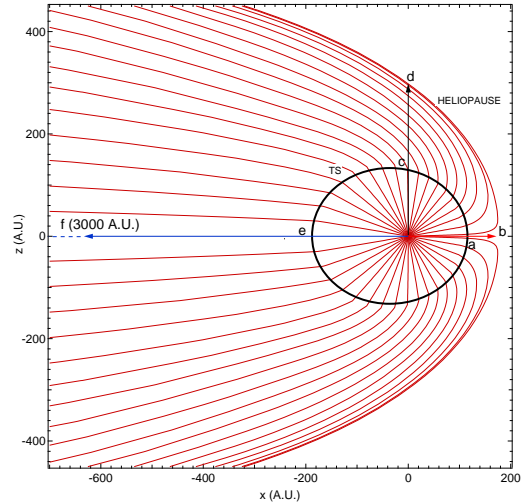}
\caption{Solar wind streamlines calculated with the Izmodenov and Alexashov (2003) model. 
Arrows mark the three LOS for which the contribution of the outer heliosphere has been calculated and presented in fig.\ref{external}. }
\label{streamlines}
\end{figure}

In Figure \ref{streamlines} are shown the solar wind streamlines from the self consistent plasma/neutrals model. The IS wind flows from right to 
left pushing the solar wind plasma and creating the heliospheric interface. In a first step, ion losses have been computed along each solar wind 
streamline. Up to the termination shock the loss term has been calculated as described earlier. Once the termination shock is crossed, we can no longer 
neglect the thermal motion of the ions which becomes important in the heliosheath, and in equation \ref{eqn:nbc} we replace $V_{SW}$ 
by $v_{rel,HS}$ where  
\begin{equation}   \label{vrel_ext}
v_{rel,HS} = \sqrt{(v_{rel}^2 + 8 k T_i/\pi m_i)}
\end{equation}
includes the relative velocity between ions and neutrals $\vec {v_{rel}} = \vec {V_{SW}} - \vec {v_n} $ and the ion thermal motion.

Except for the upwind and downwind direction, any line of sight in the heliosheath crosses all the streamlines making at the Sun an angle 
between 0 and 90 degrees. Since ions displaced from the forward direction continue to exchange with the neutral atoms the computation 
of the intensities makes use of the loss along the streamlines. For the upwind LOS emission ceases at the heliopause. For the downwind LOS ions 
contribute to the emission at large distances out to 3000 AU where the ions have all been removed by CX. 

\begin{figure}
\centering
\includegraphics[width=0.5\textwidth ,height=!]{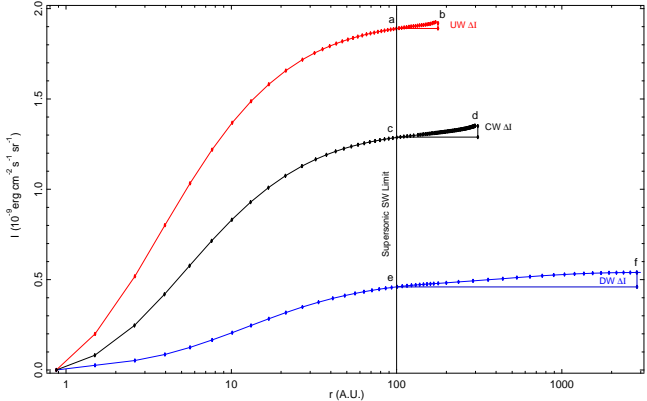}
\caption{ Contribution of the heliosheath upwind (red) and crosswind (black), 
and the heliotail (blue) to a heavy SW ion emission. Intensity, 
in units of 10$^{-9}$ erg cm$^{-2}$ s$^{-1}$ sr$^{-1}$, 
is presented as a function of distance r to the Sun (in AU). See location 
of (a) to (f) points in fig.\ref{streamlines}}
\label{external}
\end{figure}

In the heliosheath, CX with atomic hydrogen dominates. Gravitational focusing and ionization effects 
which both enhance the concentration of He over H close to the Sun are not at work at large distances and in estimating the X-rays 
from the heliosheath we may ignore He. We may also ignore solar activity which is limited in its effects at large distances and consider the intensities 
produced in a stationary slow solar wind. In figure \ref{external}, we present the total emission in units of erg cm$^{-2}$ s$^{-1}$ sr$^{-1}$ from CX 
with H as a function of distance in AU from the Sun on three LOS, upwind, crosswind and downwind, for a hypothetical ion of atomic number 16 
with a cross-section of 3.5 10$^{-15}$ cm$^{2}$ and at a spectral line of 0.56 keV.

In table \ref{External_table} we present the contributions of emissions in the outer heliosphere from individual ions, expressed as percentages 
of the total emission intensity due to H and total emission due to H + He, received within 100 AU. The corresponding amount is much larger downwind and 
depends essentially on the cross section of CX with H atoms, decreasing as the cross section increases. The largest enhancement happens for C$^{5+}$ 
which has the smallest cross section and is depleted less rapidly with distance. Upwind, the heliosheath extends to about 170 AU and its contribution 
is small at 2.7\% of the total. Crosswind there is a small increase to 7.3\% , the heliopause occuring at 300 AU. A substantial enhancement of 
38\% results for the downwind direction because CX contributes out to 3000 AU. 

As stated above, in the external heliosphere most of the emission comes from hydrogen. In Table \ref{External_table} we show the resulting 
contributions to the emission generated by collisions with H and the resulting contribution to the total emission. The contributions of the 
heliosheath to the total intensities are smaller on the downwind hemisphere since there, CX with He is the dominant source of emission. Globally, due to 
this counteracting effect, the contribution is limited to about 5\%.

\begin{table}
\caption{Contribution of the Outer Heliosphere, up to $\sim$ 170 AU UW, $\sim$ 300 AU CW and $\sim$ 3000 AU DW, to the X-ray intensity due to CX 
between heavy SW ions and IS H atoms and to the total directional X-ray intensity (H+He) within 100 AU.\label{External_table}}

\begin{flushleft}
\begin{minipage}[t]{10cm}
   \renewcommand{\footnoterule}{}
\begin{tabular}{@{}l@{}|c||c@{}|c||c|c@{}||c|c @{}}\hline \hline
\multicolumn{2}{c||}{}& \multicolumn{6}{c}{H/He emission within 100AU} \\\hline 
\multicolumn{2}{c||}{}& \multicolumn{2}{c||}{UW} & \multicolumn{2}{c||}{CW} & \multicolumn{2}{c}{DW} \\ 
\multicolumn{2}{c||}{}& \multicolumn{2}{c||}{2.7} & \multicolumn{2}{c||}{1.08} & \multicolumn{2}{c}{0.15} \\ \hline \hline
\multicolumn{2}{c||}{}& \multicolumn{6}{c}{\%  LOS} \\ \hline
$X^{Q+}$ & $\sigma_{(H, X^{Q+})}$  & H  \footnote[1]{\% to add to the emission due to CX with H }& H+He \footnote[2]{\% to add to the 
total emission due to CX with H and He}& H & H+He & H & H+He \\ 
& $(10^{-15}\,cm^2)$ & & & & & &\\ \hline
C$^{6+}$& 4.16 & 1.8 & 1.31 & 4.3 & 2.24 & 15 & 1.95 \\
C$^{5+}$& 2.00 & 2.7 & 1.97 & 7.3 & 3.8 & 37.9 & 4.93 \\
N$^{7+}$& 5.67 & 1.4 & 1.02 & 3.0 & 1.56 & 9.9 & 1.29 \\
N$^{6+}$& 3.71 & 1.9 & 1.39 & 4.7 & 2.44 & 17 & 2.21 \\
N$^{5+}$& 2.27 & 2.4 & 1.75 & 6.7 & 3.48 & 32 & 4.16\\
O$^{8+}$& 5.65 & 1.3 & 0.95 & 2.9 & 1.51 & 9.8 & 1.27\\
O$^{7+}$& 3.40 & 1.9 & 1.39 & 4.9 & 2.55 & 19 & 2.47\\
O$^{6+}$& 3.67 & 1.8 & 1.31 & 4.7 & 2.44 & 17.2 & 2.24\\
Ne$^{9+}$& 7.20 & 0.97 & 0.71 & 2.1 & 1.09 & 7.1 & 0.92\\
Ne$^{8+}$& 3.70 & 1.7 & 1.24 & 4.5 & 2.34 & 16.9 & 2.2\\
Mg$^{11+}$& 7.5 & 0.9 & 0.66 & 2.0 & 1.04 & 6.7 & 0.87\\ 
Mg$^{10+}$& 3.73 & 1.6 & 1.17 & 4.4 & 2.29 & 16.6 & 2.16\\ \hline
\end{tabular}
\end{minipage}
\end{flushleft}
\end{table}

\section{EUV and soft X-ray spectra} \label{helio_spectra}
\begin{figure}
\centering
\includegraphics[width=0.5\textwidth ,height=!]{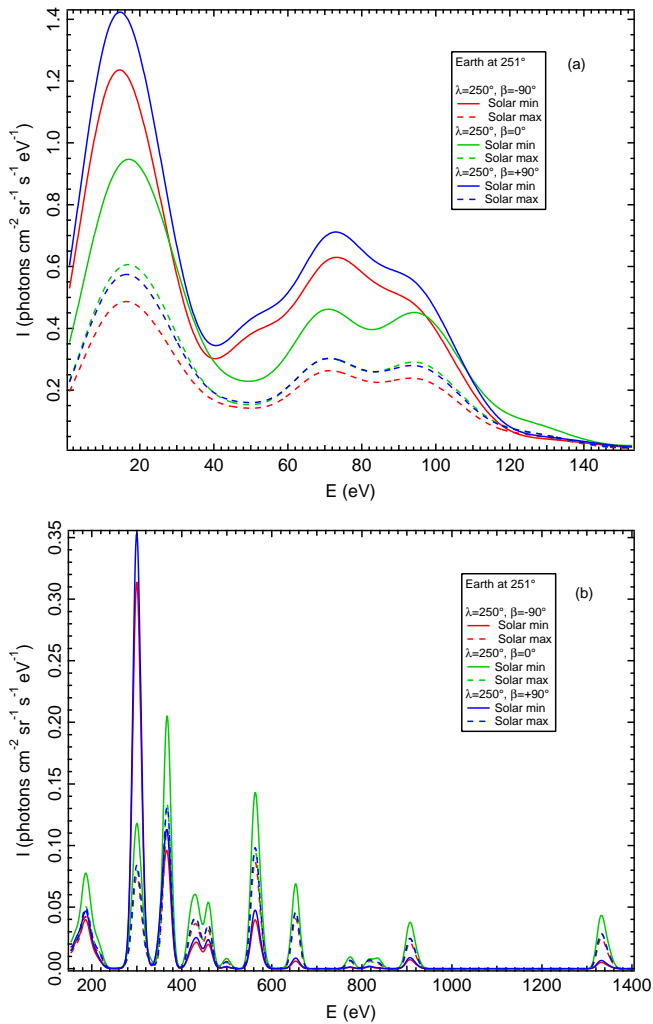}
\caption{Heliospheric EUV (a) and X-ray (b) emission spectra for an observer located upwind pointing in three different lines of sight: 
North ecliptic pole in blue, south ecliptic pole in red and equatorial antisolar LOS in green. 
Plain lines correspond to solar minimum conditions and dashed lines correspond to solar maximum conditions. 
Units are photons cm$^{-2}$ s$^{-1}$ sr$^{-1}$ eV$^{-1}$.}
\label{spectreUW}
\end{figure}

\begin{figure}
\centering
\includegraphics[width=0.5\textwidth ,height=!]{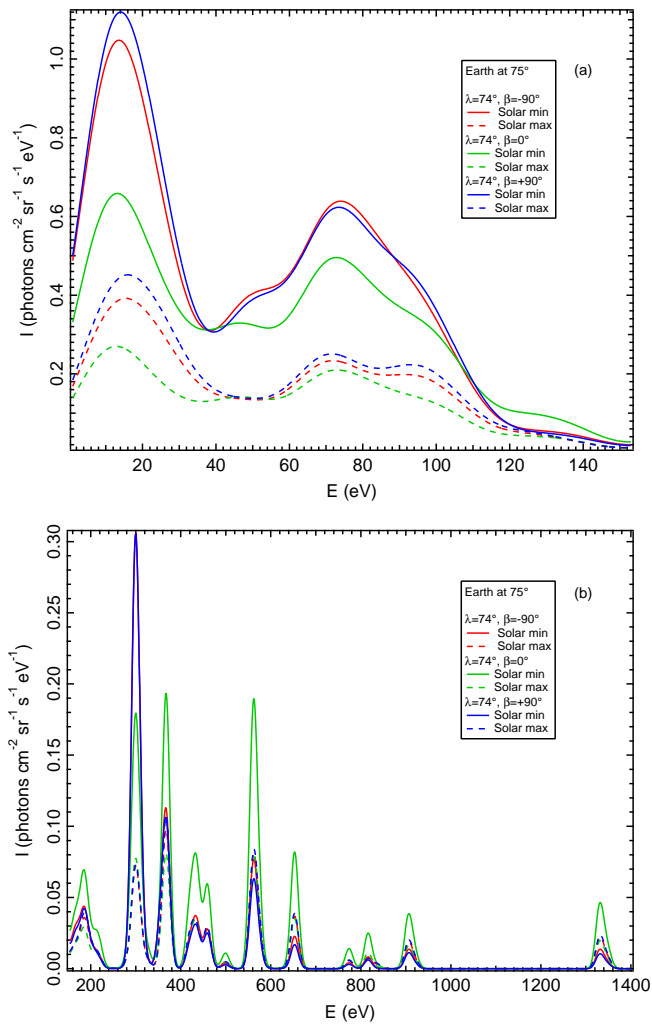}
\caption{Same as fig. \ref{spectreUW} but for an observer located downwind. Units are photons cm$^{-2}$ s$^{-1}$ sr$^{-1}$ eV$^{-1}$.}
\label{specDW}
\end{figure}

EUV and soft X-ray spectra for three LOS from two locations of the observer at solar minimum and solar maximum are presented in figures 
\ref{spectreUW} and \ref{specDW}. 

The LOS are specified by $\beta_{LOS}$ = -90\deg , South heliospheric Pole LOS, $\beta_{LOS}$ = 90\deg , the North heliospheric Pole and 
$\beta_{LOS}$ = 0\deg , in the equatorial plane. The LOS with $\beta_{LOS}$ = 0\deg is oriented towards the anti-solar direction for each 
position of the observer. Thus, when the observer is located upwind at $\lambda_E$ = 251\deg  the equatorial LOS is directed 
towards the incoming neutral atom flow, $\lambda_H$ $\sim$ 252\deg , $\beta_H$ $\sim$ 8\deg . When the observer is located downwind at 
$\lambda_E$ = 75\deg, the equatorial LOS is directed almost inside the Helium focusing cone ($\lambda_{He}$ = 74\deg , $\beta_{He}$ = -5\deg ). 

The spectra include the emission lines from the ions listed in table \ref{SWparam} , at energies between 0.005 keV to 1.4 keV. They exclude 
a small fraction of heavier ions Fe$^{Q+}$, Si$^{Q+}$ and S$^{Q+}$ which produce lines in the 300 eV energy range. Each line is a Gaussian profile 
with FWHM of 12 eV, which corresponds to the resolution achieved by McGammon et al., 2002. The EUV and X-ray spectra arising from CX vary significantly 
with solar activity, line of sight and observer location. 
    
\section{Effect of an Impulsive Solar Event on the X-ray Emission Temporal Profile: The Hubble Deep Field-North observation} \label{Snow_mod}
SCK04 observed the HDFN with the XMM-Newton European Photon Imaging Camera (EPIC) on June 1, 2001 and presented a temporal and 
spectroscopic analysis of the diffuse X-ray emission in the range $E <$ 1.5 keV. They identify in their observations the X-ray spectrum expected 
from CX emission of the highly ionized solar wind ions, characterized by C VI lines at 0.37 and 0.46 keV, O VII line at 0.57 keV 
(the triplet 0.561, 0.568, 0.574 keV), O VIII lines at 0.65 and $\sim$ 0.8 keV, Ne XI lines at $\sim$ 0.92 keV, and Mg XI lines at $\sim$ 1.35 keV. 
They observed a strong X-ray intensity enhancement in this energy range during one of the long exposures and they correlated the increase with 
an enhancement of both the solar wind flux and the highly ionized heavy ion abundances, in particular with a high O$^{7+}$/O$^{6+}$ ratio. 

They considered the possible targets to be interstellar neutral H and He atoms or exospheric H atoms of the geocorona. Neither source appears 
to provide a satisfactory explanation of the temporal variation of the intensities. The emissions and the SW enhancement cut off at the same time, 
suggesting a terrestrial origin, yet the intensity remained nearly constant over a long period of time during which the SW varied by a factor between 
3 and 6, consistent with a distributed heliospheric emission.

We explore the temporal evolution of the X-ray emission detected along a given LOS in the interplanetary space, under the influence of a solar wind 
enhancement. We model the location of the solar wind as a function of time along a Parker-type spiral, taking account of solar rotation. We assume 
that the LOS is entirely contained within the solar equatorial plane and that the rotation axis is perpendicular to the ecliptic. 
The former assumption is questionable. However, we have made some simulation in similar geometries for a 45\deg inclined LOS and for 
an active region extended in latitude, for which we obtain similar results. In addition, we are interested in the way the rotation influences 
the temporal behaviour of the emission, and how it is combined with the non-stationary ionization.

We consider two regimes of X-ray emission, high and low (Snowden et al. 2004). The observational parameters are given in table \ref{XMMparam}. 
Observations began on June 1, 2001 at 08:16:36 UT, about 0.2 d before the SW flux enhancement. The observer was at $\lambda_E$ = 251\deg , 
$\beta$ = 0.0\deg and the LOS along which XMM-Newton EPIC was pointing was $\lambda_{HDFN}$ = 148\deg , $\beta_{HDFN}$ = +57\deg corresponding 
to a heliographic latitude $\beta_{helio}$ = +50\deg . For the model the LOS is in the ecliptic plane, towards $\lambda$ = 148\deg and 
the observer is at $\lambda_E$ = 251\deg to match as well as possible the XMM-Newton observation. 

We define the proton flux in units of 2.6 10$^{8}$ cm$^{-2}$ s$^{-1}$ which is the typical SSW value for the quiet Sun at solar maximum. We 
call it the slow proton flux unit (SPFU). In our simulation the base level is 1.15 SPFU derived from ACE/SWEPAM data also (SCK04). 
From the same data we represent the SW flux during the enhancement by a step function whose area is equal to the integral of the measured flux 
during the same period of time. This corresponds to a mean value of about 3.8 SPFU, a relative increase of a factor of 3.3. The high X-ray emission 
regime occurs before and during the SW enhancement and the low X-ray emission regime starts with the SW enhancement cut-off (SCK04). 

According to SCK04 the O$^{7+}$/O$^{6+}$ ratio was significantly high at around 0.99 during the X-ray high emission regime and dropped to 
0.15 just when the SW flux measurements came back to normal values. SCK04 estimate that an equivalent enhancement should have occurred for 
the relative abundances of the ionization states of O$^{8+}$, Ne$^{9+}$ and Mg$^{11+}$, but also cite contradictory observations from preliminary 
ACE results for the O$^{8+}$/O$^{7+}$ ratio. According to these data, the O$^{8+}$/O$^{7+}$ ratio was curiously low, about 0.05, during the event and 
increased by a factor of 20 - 30 just after the event. 

We have investigated these two situations, focusing on the two oxygen ions O$^{8+}$ and O$^{7+}$, which produce excited O$^{7+}$ and O$^{6+}$ ions respectively. 
Their de-excitation gives rise to the most intense lines in the energy range 0.52 - 0.75 keV: the O VII line at 0.57 keV (the triplet 0.561, 0.568, 
0.575 keV) and the O VIII line at 0.65 keV, best detected by XMM.

\begin{table}
\caption{Parameters of XMM-Newton observation of HDFN on June 1, 2001. \label{XMMparam}}
\begin{flushleft}
\begin{minipage}[t]{10cm}
   \renewcommand{\footnoterule}{}
\begin{tabular}{@{}l|c|c|c|c@{}}\hline
\multicolumn{5}{c}{Observation geometry parameters}\\ \hline
\multicolumn{5}{c}{2001 June 01, $\lambda_{XMM}$ = 251\deg }\\
\multicolumn{5}{c}{$\lambda_{HDFN}$ = 148\deg , $\beta_{HDFN}$ = +57\deg }\\ \hline
Solar Wind Parameters & \multicolumn{2}{c}{High} & \multicolumn{2}{c}{Low}\\ \hline
H$^+$ flux SPFU \footnote[1]{ACE SWEPAM data, SPFU = 2.6 10$^{8}$ cm$^{-2}$ s$^{-1}$.} & \multicolumn{2}{c}{3.8} & \multicolumn{2}{c}{1.15}\\
O$^{7+}$/O$^{6+}$ $^b$  & \multicolumn{2}{c}{0.99} & \multicolumn{2}{c}{0.15}\\\hline
& \multicolumn{2}{c|}{Model 1 \footnote[2]{ACE SWICS data.}} & \multicolumn{2}{c}{Model 2 \footnote[3]{Implied Ion Rate SCK04}} \\
& High & Low & High & Low \\ 
O$^{8+}$/O$^{7+}$ & 0.05 ? & 1 ? & 0.57 &  \\ \hline \hline
Line & \multicolumn{4}{c}{Model Energy Flux (10$^{-9}$ erg cm$^{-2}$ sr$^{-1}$ s$^{-1}$)}\\\hline
O VII 0.57 keV & 12.5 & 7.04 & 11.5 & 7.26 \\
O VIII 0.65 keV & 6.28 & 6.5 & 6.56 & 3.38 \\ \hline \hline
\end{tabular}
\end{minipage}
\end{flushleft}
\end{table}

We have simplified the description of the solar event by considering an active region (AR) 
with a longitudinal size of about 6\deg, rotating with the Sun (360 degrees per 27 days), and assumed it persists throughout a time of 5 days. 
If we assume that the active region is continually expelling material, the enhancement will be described as a step function during about half a day. 
The expelled material, propagating radially at 400 km/s, travels for about 4.3 d before reaching the Earth while the AR turns through 57\deg 
on the solar disk. 

\begin{figure}
\centering
\includegraphics[width=0.45\textwidth , height=!]{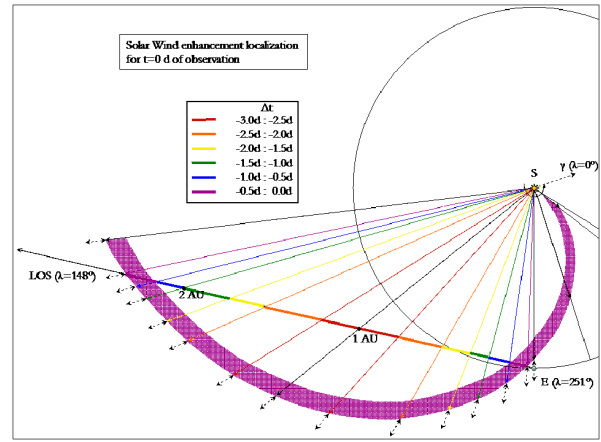}
\caption{Solar Wind enhancement localization for the start of the XMM HDFN X-ray observation.}
\label{Snowgeo}
\end{figure}

In figure \ref{Snowgeo} we illustrate the position of the SW enhancement on day 0 at the start of XMM-Newton observation. The satellite is 
at $\lambda$ = 251\deg with EPIC pointing towards $\lambda_{HDFN}$ = 148\deg 0.2 d before the measurements of the SW enhancement. The sections 
of LOS that respond to the presence of the SW enhancement, at this precise moment, are represented by the intersections of the dark purple arm with the LOS. 
All parts of the LOS between the two intersections have been reached earlier at the times indicated by the corresponding colours.

When the SW flux, enhanced by a factor E, encounters the neutral atoms in interplanetary space, the X-ray emission intensity is increased 
proportionally. In parallel, because CX with protons is the main source of ionization of IS H atoms, the ionization rate is increased 
by the same factor E. The helium ionization rate is also slightly affected. The changes may be represented approximately by the ionization rate ratios 
$\beta_{high}/ \beta_{low} = (E + 0.1)/1.1$ for H and $\beta_{high}/ \beta_{low} = (0.1\cdot E + 1)/1.1$ for He. Therefore, we have a global decrease 
in the volumic density of both neutral species which is calculated in the model. The loss of emissivity due to this decrease of neutral density, though, 
is much fainter than the direct gain from the increased SW flux. Once the SW has passed through the region the neutral densities relax to lower 
equilibrium values with a modified SW ion distribution and a diminished X-ray intensity. 

Along lines of sight, 88\% of the total X-ray emission due to H is produced in the first 10 AU and 42\% in the first 2 AU . 
For He, 98\% of the total X-ray emission is produced within the first 5 AU and 88\% in the first 2 AU. Figure \ref{Snowgeo} shows that the major 
portion of the enhanced X-ray emission on the LOS ($\lambda$ = 148\deg ) is produced \textbf{before} the SW enhancement is registered by ACE and Wind solar 
wind instruments. 

\begin{figure*}
\centering
\includegraphics[width=\textwidth ,height=!]{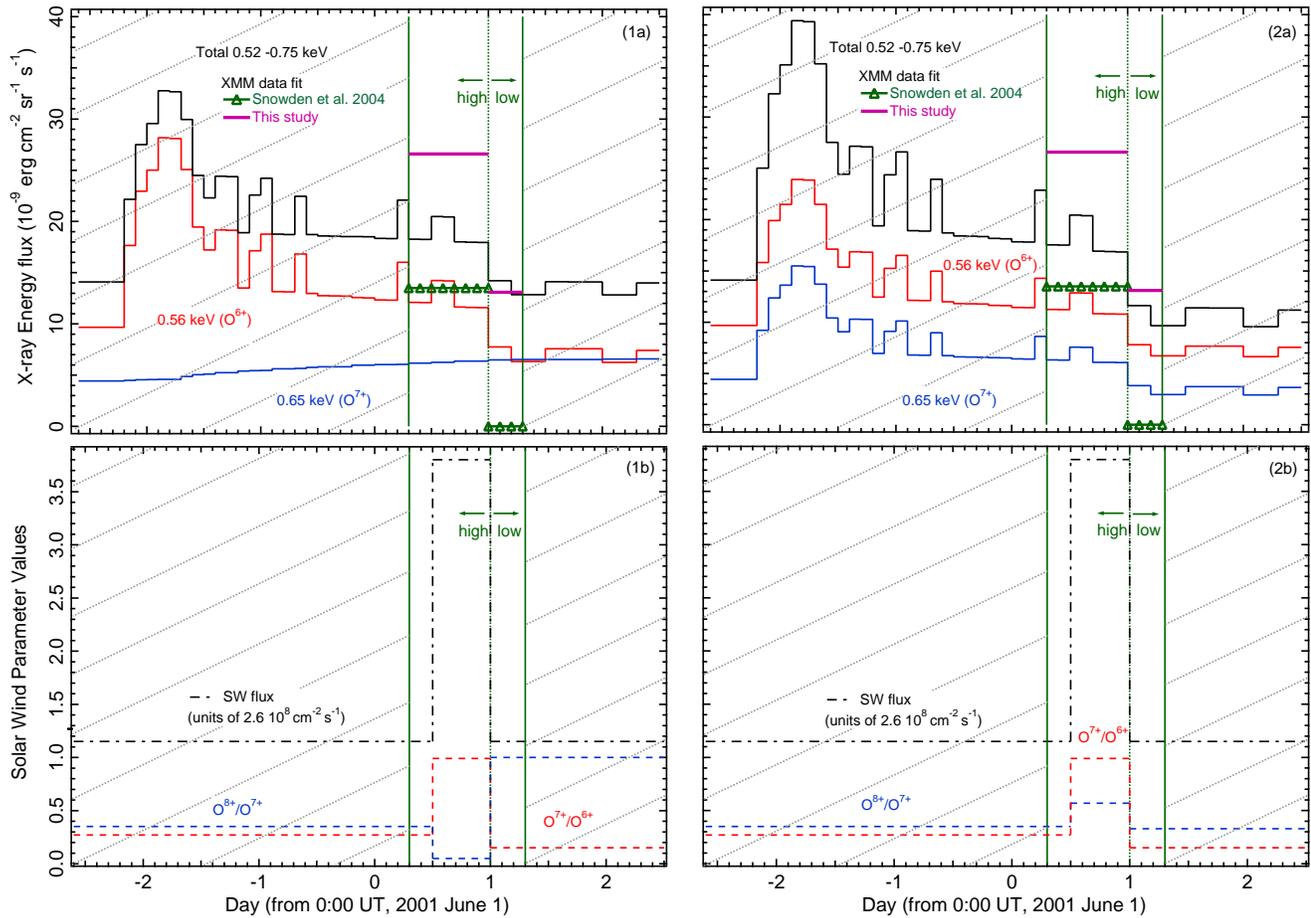}
\caption{Various parameters' temporal profiles, before, during and after the 
June 1, 2001 HDFN observation. Periods where no X-ray observations of HDFN are 
available are crosshatched in the graphs. Lower panels show SW input parameters: The dashed black line shows the proton flux 
enhancement, estimated as a step function, in units of SPFU (see text for further explanation). The dashed red line 
and the dashed blue line show the O$^{7+}$/O$^{6+}$ and the O$^{8+}$/O$^{7+}$ ratio evolution respectively. 
Left panel: O$^{7+}$/O$^{6+}$ and O$^{8+}$/O$^{7+}$ ratios derived from ACE measurements. Right panel: O$^{8+}$/O$^{7+}$ ratio 
implied by Snowden et al analysis. Upper panels show X-ray emission temporal profiles for the two cases considered:
The plain black line shows the simulation result for X-ray energy flux in the energy range 0.52 - 0.75 keV. 
The plain red and blue lines show X-ray energy fluxes for the two major 
spectral lines in this energy range: the O VII 0.56 keV and O VIII 0.65 keV lines respectively. 
The green horizontal lines show average high and low heliospheric energy fluxes in the energy range 0.52 - 0.75 keV by XMM data fit, 
assuming there is no heliospheric contribution in the low X-ray emission state (SCK04). The purple horizontal lines show 
equivalent averages assuming that there is no cosmic background contribution, only particle background (this study).
Energy fluxes are all presented in units 10$^{-9}$ ergs cm$^{-2}$ sr$^{-1}$ s$^{-1}$.}
\label{Snowtemp}
\end{figure*}

\begin{figure*}
\centering
\includegraphics[width=0.5\textwidth, height =!]{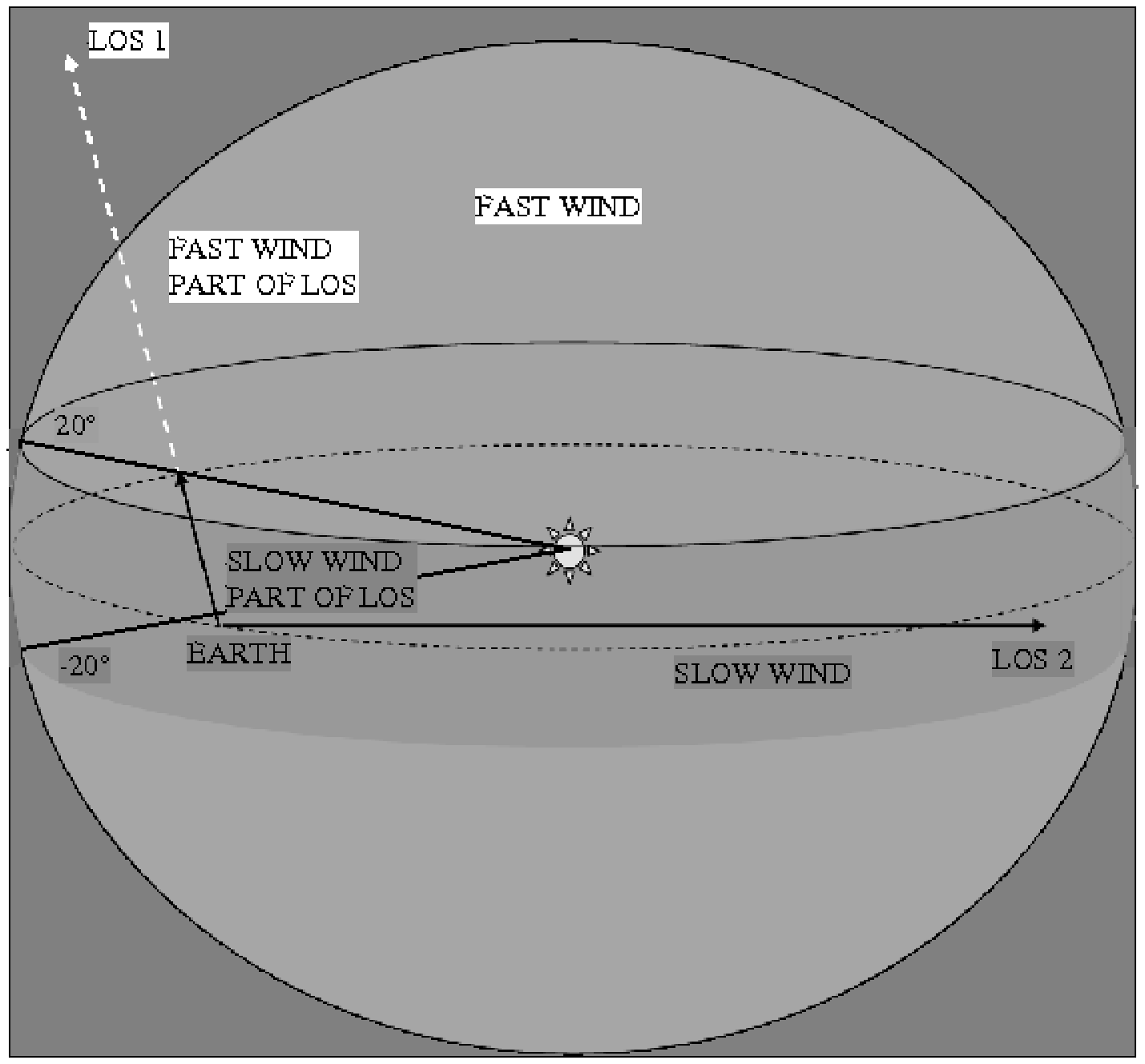}
\caption{Line of Sight (LOS) Geometry}
\label{los_geo}
\end{figure*}

\begin{figure*}
\centering
\includegraphics[width=0.6\textwidth ,height=!]{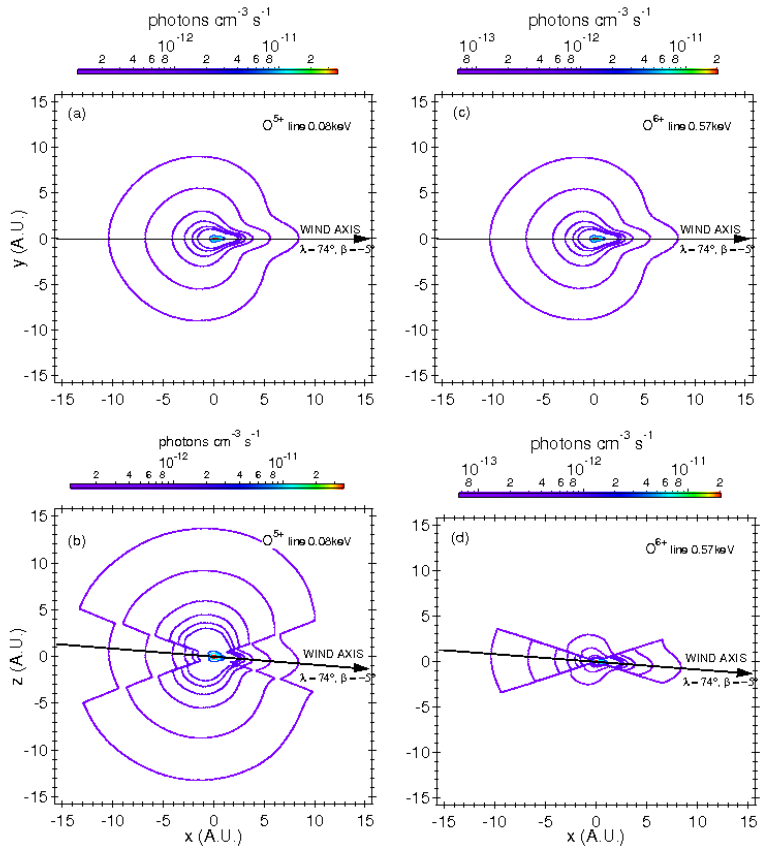}
\caption{ Iso-emissivity contour maps up to 15 AU for solar 
minimum conditions. Left column corresponds to the sum of the ion 
O$^{5+}$ lines at 0.072 and 0.082 keV and right column to the sum of the ion O$^{6+}$ lines at 0.561, 0.569 and 0.574 keV. 
Top panels correspond to the plane (x, y) defined by the vectors 
(74\deg , -5\deg ) and (-16\deg , 0\deg ). Bottom panels correspond 
to the plane (x, z) containing the IS He wind axis (74\deg , -5\deg ) 
and the ecliptic axis. The color scale is in units of photons cm$^{-3}$ s$^{-1}$. }
\label{isoemiO5_O6}
\end{figure*}

\begin{figure}
\centering
\includegraphics[width=0.5\textwidth ,height=!]{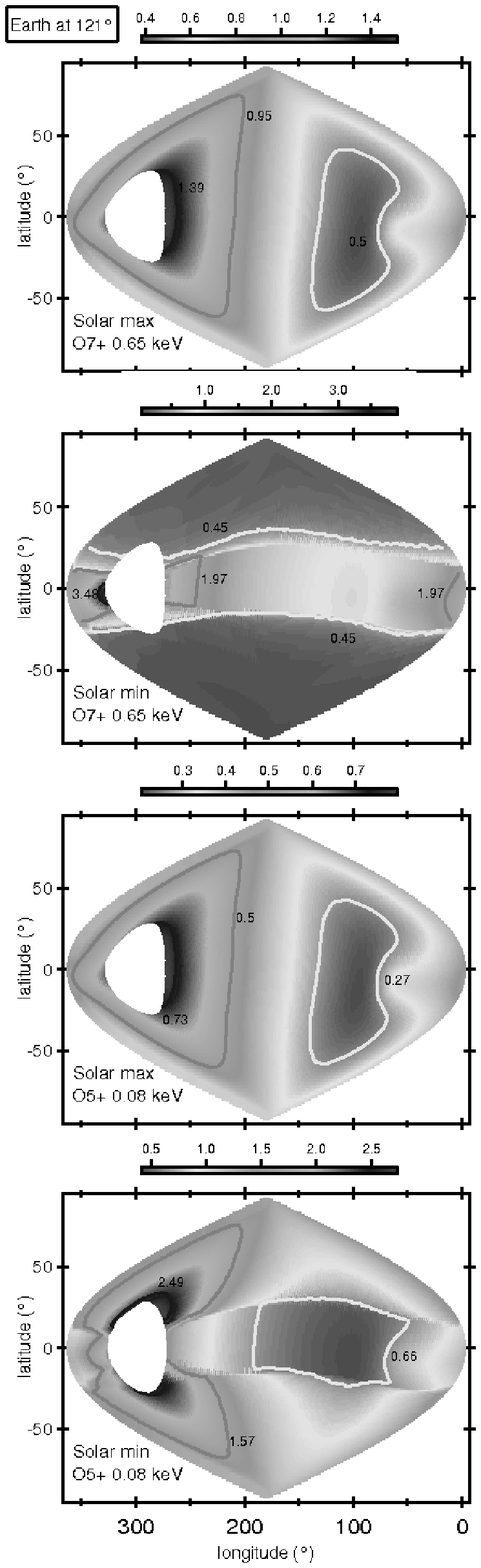}
\caption{Solar maximum and solar minimum full sky monochromatic emission maps. The two lines here are the 0.65 keV (O$^{7+}$) and the 0.08 keV 
(O$^{5+}$). The observer is situated at 121\deg ecliptic longitude. The color scale is in 
units of 10$^{-9}$ erg cm$^{-2}$ sr$^{-1}$ s$^{-1}$, red colour corresponding to minimum and blue to maximum values. 
The masked area corresponds to the 20\deg x 20\deg region around the solar disk.
The map is shown in ecliptic coordinates.}
\label{fullsky121}
\end{figure}

\begin{figure}
\centering
\includegraphics[width=0.5\textwidth ,height=!]{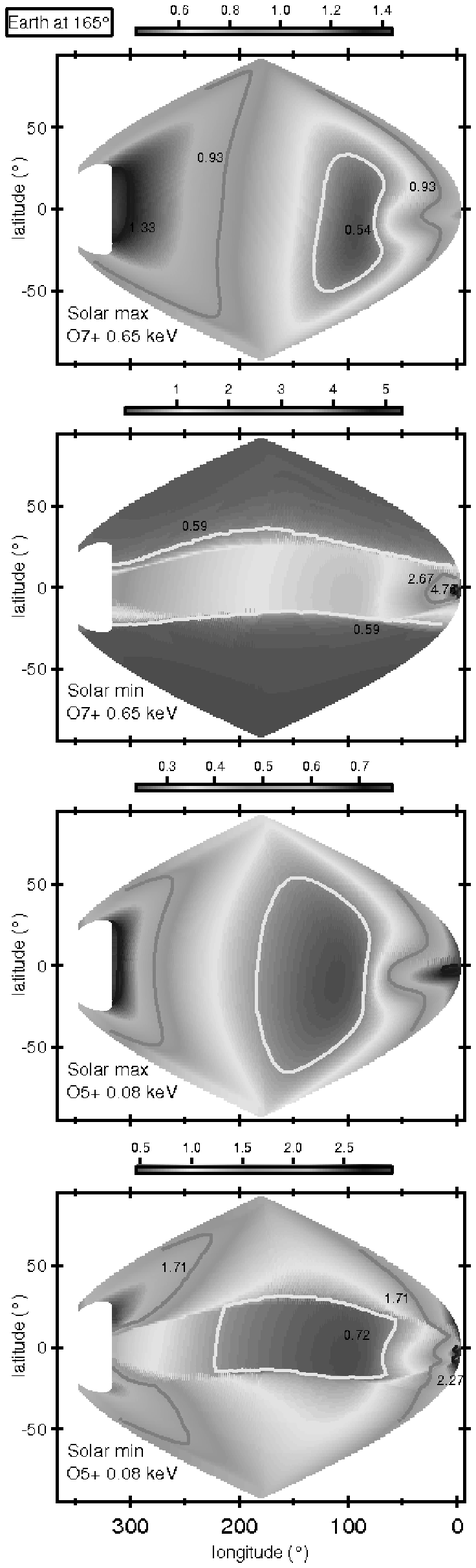}
\caption{ Same as Fig \ref{fullsky121} for an observer 
at 165\deg ecliptic longitude. The color scale is in 
units of 10$^{-9}$ erg cm$^{-2}$ sr$^{-1}$ s$^{-1}$, red colour corresponding to minimum and blue to maximum values. }
\label{fullskyCW}
\end{figure}

\begin{figure*}
\centering
\includegraphics[width=0.5\textwidth ,height=!]{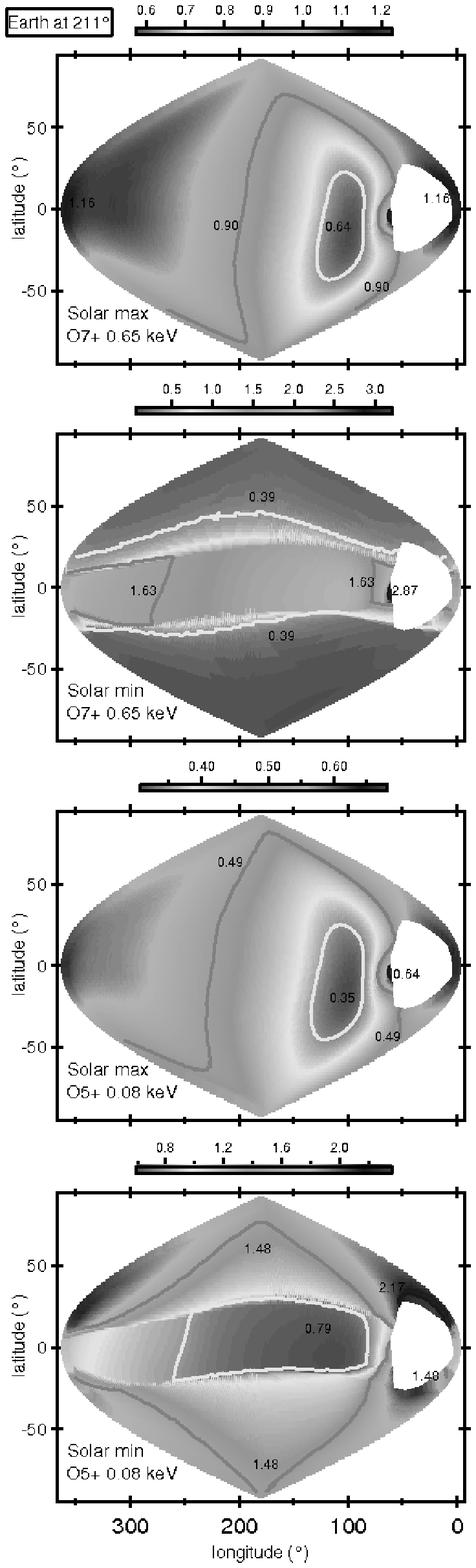}
\caption{ Same as Fig \ref{fullsky121} for an observer 
at 211\deg ecliptic longitude. The color scale is in 
units of 10$^{-9}$ erg cm$^{-2}$ sr$^{-1}$ s$^{-1}$, red colour corresponding to minimum and blue to maximum values. }
\label{fullsky211}
\end{figure*}

In figure \ref{Snowtemp} we present the results of simulations and SCK04 data. In the bottom panels we give the solar wind input parameters 
and in the upper panels the X-ray emissions derived from the simulations together with the mean values from the observational data (SCK04). 
The black dashed line shows the simulated SW flux enhancement in SPFU as it would be measured by an instrument 
at the position of the Earth ($\lambda$ = 251\deg , 1 AU). The plain green vertical lines correspond to the start and end of XMM observations, 
while the dashed green vertical line delimits the high and low X-ray emission regimes, and the end of the SW enhancement. The crosshatched parts of 
the graphs represent the periods where no X-ray observations in this LOS are available. The red and blue dashed lines represent the O$^{7+}$/O$^{6+}$ 
and O$^{8+}$/O$^{7+}$ ratios respectively, associated with the variations of SW proton flux. 

Before the SW enhancement, we adopt the ratios O$^{7+}$/O$^{6+}$ = 0.27 and O$^{8+}$/O$^{7+}$ = 0.35 (Schwadron \&\ Cravens, 2000). 
For the time during and after the SW enhancement we consider two possibilities. One, described in panels 1a and 1b, uses the ACE/SWICS data 
in the SCK04 analysis which showed a very high O$^{7+}$/O$^{6+}$ ratio during the enhancement, dropping by an order of magnitude after the event, 
and an O$^{8+}$/O$^{7+}$ ratio which was low when O$^{7+}$/O$^{6+}$ was high but increased by a factor of twenty when O$^{7+}$/O$^{6+}$ fell. 
For the second possibility, described in panels 2a and 2b, we adopt the measured O$^{7+}$/O$^{6+}$ ratio and assume O$^{8+}$/O$^{7+}$ = 0.57 
during the event as implied by the O$^{8+}$ and O$^{7+}$ lines ratio calculation in SCK04. This ratio is quite superior to the normal value (0.35) 
but not unreasonably so and moreover it involves a similar O$^{7+}$ and O$^{8+}$ evolution throughout the event. After the event cut-off we consider 
a O$^{8+}$/O$^{7+}$ ratio of 0.33 which implies again a similar decrease of O$^{7+}$ and O$^{8+}$ which is the most frequent case in solar events. 

The red and blue curves show the X-ray lightcurves for the O VII line at 0.56 keV and the O VIII line at 0.65 keV respectively. The plain black 
curve shows the simulated lightcurve for total X-ray emission in the 0.52-0.75 keV energy range. 

Figure \ref{Snowgeo}, demonstrates that the LOS is affected by the SW enhancement about 3.0 d before the begining of observations, 
which corresponds to around day -2.6 on our lightcurve. The X-ray emissions contained in the two lines evolve in very similar ways in the two models. 
Calculated with ACE data (left panel), it rises suddenly by about 130\% half a day later, to reach a maximum on day -1.8 from June 1, 0:00 UT. Then, 
a first drop occurs on day -1.0 at $\sim$ 57\% of the maximum value and a second at $\sim$ 40\% of the maximum value on day +1. This is exactly 
the moment the SW enhancement ends. The emission calculated with the O$^{8+}$/O$^{7+}$ ratio (right panel) implied by the spectral lines rises and 
falls at exactly the same intervals and more steeply. The emission rises by $\sim$ 150\% , then decreases by 53\% and at the last cut-off, 
on day +1, is stabilized  at 30\% of the maximum value. 

The main difference between the two possible scenarios lies in the evolution of the two OVII and OVIII lines. With the second set of conditions, 
the lines behave in similar ways, decreasing in intensity by factors 1.6 and 1.94 respectively. The mean high intensities are 
11.5 10$^{-9}$ erg cm$^{-2}$ sr$^{-1}$ s$^{-1}$ for O VII and 6.5 10$^{-9}$ erg cm$^{-2}$ sr$^{-1}$ s$^{-1}$ for O VIII, and the mean low 
7.26 10$^{-9}$ erg cm$^{-2}$ sr$^{-1}$ s$^{-1}$ for O VII and 3.38 10$^{-9}$ erg cm$^{-2}$ sr$^{-1}$ s$^{-1}$ for O VIII. According to the first scenario 
the OVIII line is nearly constant varying between 6.28 10$^{-9}$ erg cm$^{-2}$ sr$^{-1}$ s$^{-1}$ and 6.5 10$^{-9}$ erg cm$^{-2}$ sr$^{-1}$ s$^{-1}$), 
and the OVII line decreases strongly from 12.5 10$^{-9}$ erg cm$^{-2}$ sr$^{-1}$ s$^{-1}$ to 7.04 10$^{-9}$ erg cm$^{-2}$ sr$^{-1}$ s$^{-1}$. 

Because there are no observations of the HDFN before day 0.3 after June 1, 0:00 UT, our comparison with the data is limited to the interval 
0.3 : 1.3 day from June 1, where the second cut-off occurs. The simulated lightcurves at the beginning of XMM observation are already at a high mean level, 
and present faint variations until the moment the SW enhancement cuts-off, on day +1. This is exactly the temporal pattern of the 0.52-0.75 keV 
emission recorded by XMM, as can be seen in Fig 1 of Snowden et al. This emission drops by a factor of about 1.5. 

To proceed further requires additional assumptions on the heliospheric and cosmic background radiations for this direction. We consider 
two extreme cases.

In the first case we assume, following Snowden et al, that the high emission level is significantly contaminated by the heliosphere while during the 
low level period this local emission is negligible and most of the signal is of cosmic origin (except for the particle background). The difference 
between the two high and low spectra provides a measurement of the heliospheric contribution during the event. Since the heliospheric 
contribution is contained in lines only, and is made of the OVII + OVIII lines in the 0.52-0.75 keV range, it  amounts to 13.5 10$^{-9}$ ergs cm$^{-2}$ 
sr$^{-1}$ s$^{-1}$ as derived from line fitting by SCK04 (Table 2, SCK04, sum of the two O lines). This absolute level is 
significantly smaller than the model estimates of about 18.8 10$^{-9}$ erg cm$^{-2}$ sr$^{-1}$ s$^{-1}$ and 
18.0 10$^{-9}$ erg cm$^{-2}$ sr$^{-1}$ s$^{-1}$ for the first and second case respectively. In contrast, the low level is effectively zero. 
We have plotted in fig. \ref{Snowtemp} the two values (13.5 and 0 $\times$ 10$^{-9}$ erg cm$^{-2}$ sr$^{-1}$ s$^{-1}$). Despite the fluctuations 
on the measured intensities they are inconsistent with the simulations, which imply a relative decrease by less than a factor of 2 
(more precisely 1.4 and 1.7 respectively). 

If we assume that there is no signal other than from the heliosphere (except for the particle background) we can use the background subtracted high and low 
XMM spectra, derived from Fig 3 of SCK04, to estimate the heliospheric contributions. We obtain high to low ratios of 1.66 for the OVII line and 2.28 
for the OVIII line. These ratios match those obtained in the second scenario but the total high intensity is 25 10$^{-9}$ erg cm$^{-2}$ sr$^{-1}$ s$^{-1}$, 
much higher than the model predictions. However in view of the many simplifying assumptions, we cannot conclusively exclude a model with a 
very low cosmic background.

Our conclusion here is that very likely the actual situation is somewhere between the two extreme cases. Still, the good agreement on the light curves 
and the overall intensity for model and data in the second case strongly suggests that (i) heliospheric X-rays are the main contributor to the 
observed enhancement and (ii) even the LOW spectrum contains a large fraction of heliospheric emission.

\section{Discussion} \label{conclusions}
We have performed a parametric study of the heliospheric X-ray background emission and an analysis of the multiple factors which can influence the 
emission level. We have demonstrated the complex response to these factors, which include the date of observation (through the difference between the 
earth longitude and the interstellar wind axis longitude), the solar cycle phase (through the H and He densities which vary significantly with the 
phase), the solar wind type (through the high ion absolute and relative abundances), and finally the solar wind history (through the potentially very 
strong solar wind enhancements and abundance variations). The contribution to the signal generated in the inner heliosphere by secondary collisions has 
been shown to be negligible. The contribution to the signal of the external heliosphere (heliosheath and tail) is estimated and found to be small, 
except in the tail direction for some particular ions. This type of model results should allow an estimate of the expected range for the emission 
for a given direction of sight and date. 

We have modelled in a very simplified way the temporal variation of the signal in the case of a sudden and temporary solar wind increase, and made a 
data/model comparison in the case of the XMM-HDFN.  It is possible to reproduce the intensity level and the temporal evolution of the line intensities, 
although no solution has been found here matching perfectly all the data simultaneously. It is likely however from this study that a large fraction of 
the signal is of heliospheric origin, including in the post-event (LOW) period of time. This is in agreement with the results of Cravens et al (2001), 
based on correlations with the solar wind, and of Lallement (2004) based on the Local Bubble emission pattern. 

It is clear from figure \ref{Snowgeo} that each specific geometry will produce a different temporal variation. A LOS at 180 degrees from the HDFN 
direction would have had a totally different light curve, for the same solar wind history. Such complex temporal variations preclude a good correlation 
on small time scales between the X-ray emission and the locally measured solar wind. On the other hand, this diversity can be useful because the 
combination of numerous examples such as the above one may allow constraints on the actual local and cosmic contributions.

To reach this goal a number of improvements in the modelling have to be worked out. We have been using here a single fluid model for the IS H 
density, but we know that H atoms are separated in the outer heliosphere in two populations with different temperature, velocity and arrival direction 
(e.g. Lallement et al., 2005). Two population models should be developed. More important, in our simulation the eruptive region is stationary in the 
Sun's frame, like a garden hose, and apart from one enhanced stream we have only used simplified stationary states of solar cycle minimum and maximum 
conditions.  Realistic models should include 3D solar wind hourly data (ideally from different vantage points) as well as precise measurements of the 
high ion relative abundances. The exact 3D observation geometry should be, as well, taken into account in the model, which requires the knowledge 
of the active region geometry. In parallel, a more precise time-dependent model of the neutral distributions, consistent with the 
solar wind flux and velocity variations is needed.\\\\ \small{\textit{Acknowledgements.} We wish to thank John Raymond for useful discussions and 
valuable suggestions which led to the correction of a mistake.\\
R.L. and D.K., V.I. acknowledge funding by CNRS and RFBR respectively under PICS contract 3205.\\
A.D. and V.K. acknowledge NASA for support through grant NNG04GD57G.}

\Online

\end{document}